# Reversal-mechanism of in-plane current induced perpendicular switching in a single ferromagnetic layer

Chong Bi and Ming Liu

Laboratory of Nano-fabrication and Novel Devices Integrated Technology, Institute of Microelectronics, Chinese Academy of Sciences, Beijing 100029, China.

E-mail: cbi@email.arizona.edu and liuming@ime.ac.cn

Abstract

We propose a magnetization reversal model to explain the perpendicular switching of a single ferromagnetic layer induced by an in-plane current. Contrary to previously proposed reversal mechanisms that such magnetic switching is directly from the Rashba or spin Hall effects, we suggest that this type of switching arises from the current-induced chirality dependent domain wall motion. By measuring the field dependent switching behaviors, we show that such switching can also be achieved between any two multidomain states, and all of these switching behaviors can be well explained by this model. This model indicates that the spin Hall angle in such structures may be overestimated and also predicts similar switching behaviors in other ferromagnetic structures with chiral domain walls or skyrmions.

**I. Introduction**

Spin-orbit (SO) torques arising from the spin Hall effects (SHE) and Rashba effect have attracted considerable attentions recently because they provide an alternative way to switch magnetization as well as drive domain wall (DW) motions[1–5]. Compared with conventional spin transfer torques (STTs), they do not need another ferromagnetic layer as a spin polarizer and the high driving current density does not need to cross the tunnel barrier[6,7]. However, there are many debates on the SO torque induced magnetization dynamics, for example, whether the spin Hall torque (SHT) from SHE [4,5] or Rashba effect[3] is the dominant source to induce the magnetic switching. Even for the basic magnetic switching mechanism, it is still not well established. It was first assumed that there was an equivalent effective perpendicular field related with the in-plane Rashba field [3], and then suggested that there should be a large SHT from the adjacent heavy metal layer [4,5], driving the magnetic switching. Particularly for the SHT model, the authors have built a macrospin model by considering SHT and all possible field torques [5], and have also observed the opposite switching directions by changing the sign of spin Hall angle [4,5]. However, both the current-induced perpendicular and collinear in-plane effective fields can be observed in such structures [8,9], which cannot be explained solely by SHT. On the other hand, the chirality dependent DW motion shows that the interfacial Dzyaloshinskii-Moriya interaction (DMI) also plays a crucial role in the current induced magnetization dynamics [1,2]. Furthermore, the macrospin model cannot explain why the critical current density for magnetization reversal ($J_c$) strongly depends on the DW pinning strength [10].

In this paper, we propose another reversal mechanism of the SO torque induced perpendicular switching based on the current induced chirality dependent DW motion. We suggest that the injected in-plane current simultaneously nucleates domains and drives DW motions in these structures, and the applied in-plane field plays the role to selectively modulate DW motions. We have experimentally demonstrated that this model can fully explain all the external field dependent switching behaviors and have revealed the relationship between $J_c$ and DW motion.

**II. Perpendicular switching model based on current-induced domain nucleation and expansion**

There are three essential conditions for this model: (1) a sufficiently large current that can induce domain nucleation and finally drive magnetization to a stable equilibrium state with 50% up (↑) and down (↓) domains; (2) the sufficiently large current can drive DW motion; (3) the up-down (↑↓) DW and down-up (↓↑) DW motions can be separately modulated by an applied external field. Figure 1 schematically illustrates this model. For an initial uniformly magnetized state (Fig. 1(a)), an injected current breaks the uniform state by nucleating random domains (Fig. 1(b)). When the current is large enough, the magnetization achieves a steady multidomain state with equal ↑ and ↓ domain areas, and meanwhile, the current drives DWs moving at the same velocity (Fig. 1(c)). When applying an external field, one type DW (↑↓ or ↓↑) motion is promoted and another type DW motion is suppressed, depending on the current and field directions (Fig. 1(d) and (e)). Therefore, the steady multidomain state with equal ↑ and ↓ domain areas will be broken and the total magnetization exhibits an orientation preference. With increasing the external field, the velocity difference between two type (↑↓ or ↓↑) DWs becomes more pronounced, and one orientation domains are expanded and another orientation domains are contracted. Because the large applied current favors to form a 50% ↑ and ↓ domain state as required by condition (1), a reversal domain will nucleate immediately within an expanded domain to keep equal ↑ and ↓ domain areas once the expanded domain area exceeds a critical value. However, as shown in Part IV, if the velocity difference becomes large enough that the reversal domain cannot nucleate, it will be expected one orientation domains are completely annihilated and the magnetization is under a dynamic uniformly magnetized state. In this case, the magnetization can be easily reversed by changing the current or external field directions. Hereafter, we will show that the three conditions can be fully satisfied in heavy metal/ferromagnet/oxide (HM/FM/MO) multilayer structures.

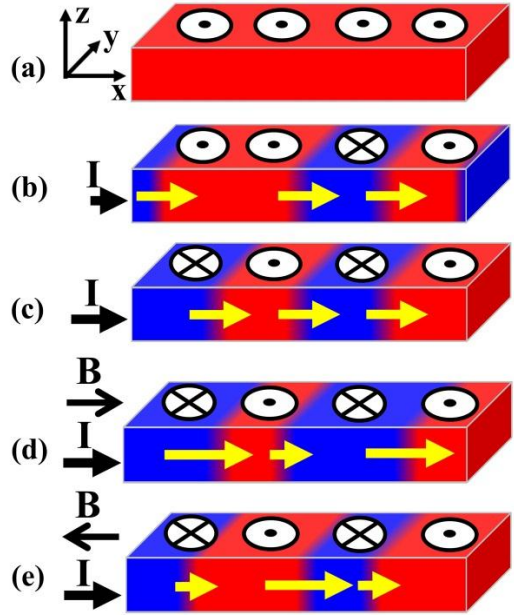

FIG.1. Schematic illustration of perpendicular switching mechanism based on current-induced domain nucleation and expansion. (a) A uniformly magnetized initial state. (b) An injected current induces some random domains within the ferromagnet and simultaneously drives DW motions. (c) Under a sufficiently large current, the random domain nucleation state evolves to a stable 'dynamic equilibrium state' with 50% ↑ and ↓ domains. (c, d) An in-plane field collinear with current flow direction breaks the 50% ↑ and ↓ domain state by modulating ↑↓ and ↓↑ DW velocities, and the domains are expanded or contracted depending on the current and field directions. The arrow on the DWs indicates its velocity.

For the first condition, as shown below, we experimentally demonstrate that, without any external fields, a large current can break a uniform magnetization state by forming 50% ↑ and ↓ domains in HM/FM structures. Generally, the current-induced domain nucleation can be caused by Joule heating and current-induced spin torques[11–14]. However, by monitoring the sample resistance change during the applied current pulses, we show that the current-induced temperature rise is only several K, which cannot dominate the domain nucleation process. On the other hand, as shown below, the thermal effect cannot explain the current direction dependent results. Therefore, we tend to believe that the spin torques dominate the domain nucleation process. Early STT theories have shown that a large current always induces the instability of a uniform magnetization, and the instability will trigger reversal domain nucleation[11–14]. Consequently, under the large current, if a uniformly magnetized area is larger than a critical value, there will be a reversal domain formed immediately to stabilize the magnetization, resulting in a steady multidomain state with equal ↑ and ↓ domains. In a normal FM, the current-induced magnetization instability arises from conventional STTs[11–14], but in HM/FM/MO structures, SHT and DMI must also be involved in the domain nucleation process. The investigation of the detailed domain nucleation mechanism in these HM/FM/MO structures is beyond the scope of this paper. However, as demonstrated by the following experimental results, one thing is clear that the large current can eventually make the magnetization evolve to a stable 'dynamic equilibrium state' with equal ↑ and ↓ domain areas, which is the crucial process for the following controllable magnetization behaviors. It is called a 'dynamic equilibrium state', because once the uniformly magnetized area (a domain) becomes larger than a critical value, a reversal domain will be immediately formed within the domain to keep the same ↑ and ↓ domain areas. As mentioned above, this occurs because equal ↑ and ↓ domain areas can stabilize the magnetization under a large current. In previous works, there is no significant domain nucleation process observed in these structures when investigating the current driven DW motion[1,2], thus the critical current for domain nucleation must be larger than that for driving DW motion, which indicates that, once a domain forms, the DWs must be moved immediately driven by the injected current. This is for condition (2).

For condition (3), it has also shown that an external field collinear with current flow direction can modulate ↑↓ and ↓↑ DW motions oppositely in FM/HM structures [1,2]. This is due to the unique chirality dependent DW motions in these structures, which can be explained by the combination of SHT and DMI [1,2]. As we expected, the external field increases one type (↑↓ or ↓↑) DW velocity and decreases or even reverses another type DW velocity. As mentioned above, the selectively control behaviors must lead to one type domain (↑ or ↓) expanded and another type domain (↓ or ↑) contracted. On the other hand, the DW motion in these structures has much higher and even opposite velocity compared with that in usual FM [1,2,15,16]. This high DW velocity provides the possibility to completely annihilate one orientation domains during the current-induced domain motion by applying a modulation field. It should be mentioned that if the applied modulation field cannot solely change magnetization states, the dynamic magnetization state will be non-volatile after removal of the injected current.

## III. Experimental results

To verify this model, we fabricated typical Pt(2.5)/Co(0.6)/AlO$_x$(1.6) structures by dc magnetron sputtering and plasma oxidation, where the number inside parentheses is the thickness in unit of nm. The samples were obtained by plasma oxidizing Pt(2.5)/Co(0.6)/Al(1.6) multilayers deposited on Si/SiO$_2$ substrates. After plasma oxidation, the samples were annealed in high vacuum at 493 K for 40 mins. The samples were then patterned into Hall bar structures with a width of 2.5 µm. After that, the Ta(5)/Cu(200) electrodes were deposited for electrical measurements. The magnetization states were detected by the anomalous Hall effect. The anomalous Hall resistance ($R_H$), which is proportional to the net perpendicular magnetization of FM ($M_z$), was measured by a 50 µA constant current. The applied current was injected by a 20 µs current pulse with a tunable amplitude, and the real injected current ($I_p$) was also measured simultaneously. The experimental setup is shown in the inset of Fig. 2(a). The positive $I_p$ corresponds to the current direction along +$x$ direction. In all measurements, we first injected a 20 µs current pulse and then applied a 50 µA dc current immediately for measuring $R_H$. We define $\eta_{\uparrow(\downarrow)}=S_{\uparrow(\downarrow)}/(S_\downarrow+ S_\uparrow)$ as the ratio of ↑(↓) domains, where $S_{\uparrow(\downarrow)}$ is the total area of ↑(↓) domains in ***xy*** plane. Therefore, $R_H$ is proportional to ($\eta_\uparrow$-$\eta_\downarrow$). All of the measurements were performed at room temperature.

We confirmed the sample with perpendicular magnetic anisotropy (PMA) by measuring $R_H$ under a perpendicular magnetic field ($B_z$). Figure 2(a) shows the measured $R_H$ as a function of $B_z$. The 100% remanence indicates that the uniformly magnetized state was retained after removal of $B_z$, confirming the existence of PMA. To demonstrate the applied current can induce the 50% ↑ and ↓ domain states, we first initialized the magnetization to uniformly magnetized ↑ or ↓ states by a perpendicular field, and then injected a series of current pulses after removal of the perpendicular field. After each pulse injection, $R_H$ was measured immediately. The amplitude of injected current pulse was gradually increased until a constant $R_H$ value was achieved, and then was decreased, as indicated by arrows in Fig. 2(b). Figure 2(b) shows the measured $R_H$ as a function of $I_p$. When -4.24 mA < $I_p$ < 4.24 mA, no significant change of $R_H$ is observed. With increasing $I_p$, the value of $R_H$ begins to decrease, and finally reaches a constant value near 0 Ω when the value of $I_p$ exceeds the critical value of 6.95 mA. The stable $R_H$ then cannot be changed whether increasing or decreasing $I_p$. All of these measurement results can be remarkably reproducible, which excludes the possibility that the sample is broken by the applied current and indicates that the decreasing $R_H$ is due to the domain nucleation. We choose $I_c$ = 4.24 mA as the critical current for domain nucleation. The final stable state with $R_H \approx 0$ Ω indicates that the film was filled by 50% ↑ and ↓ domains under the large current injection, which is consistent with our expectation. The corresponding current density of $I_c$ is about $5.47 \times 10^{11}$ A/m$^2$, which is much larger than the critical current for depinning DWs in similar structures [1,2], further confirming that the injected current not only induces domain nucleation but also simultaneously drives DW motion. Because all DWs have the same velocity without external field [1,2], the formed domains are under the 'dynamic equilibrium state' with $\eta_{\uparrow(\downarrow)}$ = 50% when $I_p$ is larger than 6.95 mA.

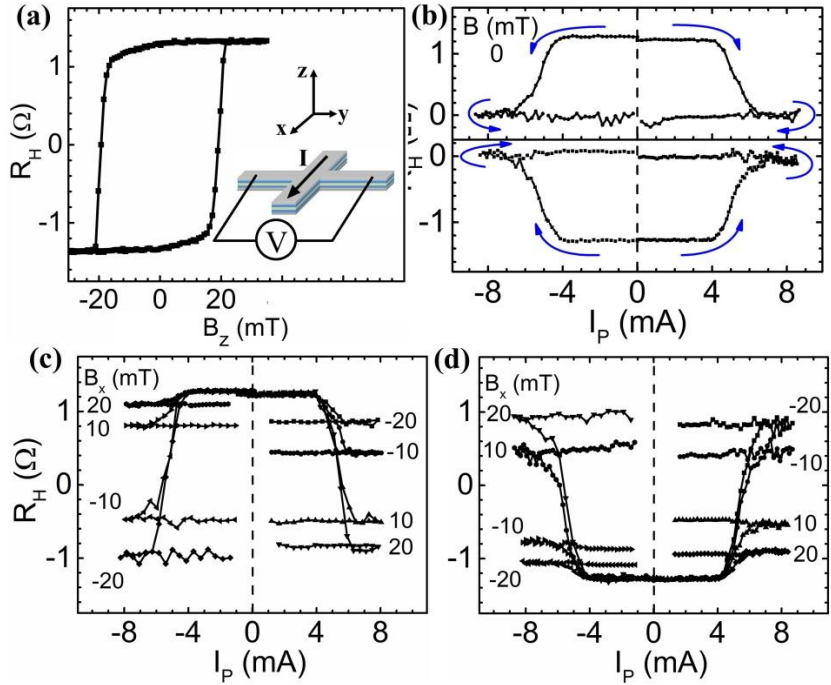

FIG.2. (a) $R_H$ as a function of perpendicular field. The inset schematically illustrates the measurement setup. (b) The measured $R_H$ after each current pulse injection as a function of $I_p$ without external fields. The magnetization was first initialized to ↑ (top panel) or ↓ (bottom panel) state by a perpendicular field before each measurement. The amplitude of injected pulse was first increased until a constant $R_H$ was achieved, and then was decreased, as indicated by arrows. (c, d) The measured $R_H$ after each current pulse injection under different $B_x$ with initialized (c) ↑ or (d) ↓ states.

It is shown that the DWs in these structures are the Néel type DWs with left-hand chirality, and an applied external field collinear with current flow direction has opposite effects on the ↑↓ and ↓↑ DW motions [1,2]. For a positive current, the applied field at the +x direction ($B_x$) will suppress ↑↓ DW motion and promote ↓↑ DW motion, and thus a positive $B_x$ will expand ↓ domain and contract ↑ domain as mentioned above. The expansion and contraction effects are opposite either for reversal current or field directions. To demonstrate the expansion and contraction effects, we measured the current induced domain nucleation under $B_x$. We also first applied a perpendicular field to initialize the magnetization and then measured $R_H$ after each current pulse injection under $B_x$. Fig. 2(c) and (d) show the measured $R_H$ as a function of $I_p$ under different $B_x$ with initialized ↑ and ↓ states, respectively. Because of the strong PMA of our sample, there is no significant magnetization tilt induced by $B_x$ up to 20 mT, as shown in Fig. 2(c) and (d), when the value of $I_p$ is smaller than about 4 mA. Similar to Fig. 2(b), the increasing injected current also induces the domain nucleation, but the final stable $R_H$ varies dramatically. It now stabilizes at a positive or negative value, instead of 0 Ω as shown in Fig. 2(b), depending on the applied field strength and direction. The final stable $R_H$ does not depend on the initialized magnetization state and is opposite for either reversal field or reversal current flow direction. Whether for ↑ or ↓ initialization states, the same final states are obtained for the same current and field. We also measured the current induced domain nucleation from other initialization states and observed the same results.

Fig. 3(a) shows the final stable $R_H$ as a function of $B_x$. The stable $R_H$ was measured after injecting a current pulse whose amplitude is larger than the critical value of 6.95 mA at each $B_x$.

$B_x$ was changed from -140 mT to 140 mT with a step of 5 mT and then back to -140 mT. Fig. 3(a) presents four curves when $I_p = \pm 7.73$ mA and $\pm 8.72$ mA. No hysteresis-like behavior is observed in these $R_H$ versus $B_x$ loops, further confirming that the final stable magnetization state does not depend on the initial sates. No difference between $I_p = \pm 7.73$ mA and $\pm 8.72$ mA loops confirms that the final stable states cannot be changed even by injecting a larger current, and also indicates that the Joule heating related thermal effects can be ignored. As mentioned above, the ignorable thermal effects were also confirmed by monitoring the sample resistance increase during current pulse duration. The final stable $R_H$ gradually changes near 0 mT and finally approaches the saturation value when $B_x$ is larger than 40 mT. All of these results are completely consistent with our model and support the case of left-hand chiral DW motions in Ref. 1. As mentioned in Ref. 1, for the structures with left-hand chiral DWs, a positive $B_x$ will expand ↓ domains for a positive current and expand ↑ domains for a negative current. For a negative $B_x$, the orientation of expanded domains will be reversed. These current induced domain behaviors under $B_x$ are clearly presented in Fig. 3(a). The gradually changing $R_H$ near 0 mT is due to the gradually changing velocity difference between ↑↓ and ↓↑ DWs induced by $B_x$. When $B_x$ is larger than 40 mT, the velocity difference becomes large enough that the reversal domains cannot nucleate, and thus the magnetization reaches a dynamic uniformly magnetized state. A little decrease of $R_H$ value in high field region is due to the in-plane field induced magnetization tilt.

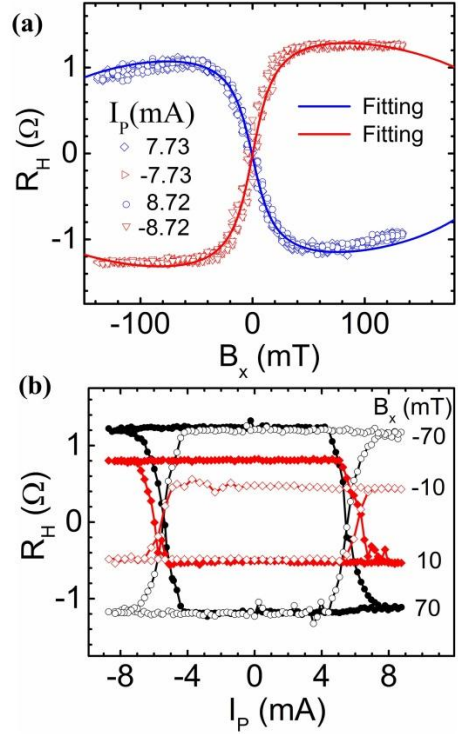

FIG.3. (a) The measured $R_H$ after a current pulse injection as a function of applied $B_x$ when $I_p = \pm 7.73$ mA and $\pm 8.72$ mA. The solid lines are the fitting results by Eq. (6). (b) The measured $R_H$ after each current pulse injection as a function of $I_p$ under a fixed $B_x$.

Because the final stable state does not depend on the initial sates, the magnetization can be reliably operated between any two stable states by controlling the current flow directions and external fields. Fig. 3(b) shows the magnetization switching between two stable states by sweeping $I_p$ at a fixed $B_x$. $R_H$ was also measured after each current pulse. For $B_x = \pm 10$ mT, the switching is only between two multidomain states, while for $B_x = \pm 70$ mT, the magnetization can be switched between uniformly magnetized ↑ and ↓ states. The field and current flow direction dependences of the magnetization switching consist with previously reported perpendicular switching [3–5], and the stable $R_H$ values under large current injections are also the same as those in Fig. 3(a).

**IV. One-dimensional model for perpendicular switching**

To get a more quantitative description of the current-induced perpendicular switching, we develop a simple one-dimensional analytical model by considering SHT and DMI effective fields. As mentioned above, without external fields, a sufficient large current favors to form the 50% ↑ and ↓ domain state, and if there is a domain larger than a critical value, a reversal domain will nucleate immediately within the domain. Considering a one-dimensional model, we assume that the critical length (along the wire long axis) that a domain can keep stable under the large current is $L_0$ and the reversal domain nucleation time is $\tau$. In absent of an external field, all ↑↓ and ↓↑ DWs are driven by the current to move at the same velocity and $\eta_{\uparrow(\downarrow)}$ keeps at 50%. When $B_x$ is applied, the velocities of ↑↓ and ↓↑ DWs will be altered oppositely, and one orientation domains will be expanded and another orientation domains will be contracted. If we assume $B_x$ does not influence the reversal domain nucleation process, the largest length of expanded domain will be $L_B = L_0 + \Delta v_B \tau$, where $\Delta v_B$ is the velocity difference between ↑↓ and ↓↑ DWs, and $R_H \propto (\eta_\uparrow - \eta_\downarrow) = 2\Delta v_B \tau / L_0$.

We use the well-established one-dimensional current-induced DW motion model [2] to investigate the $B_x$ induced $\Delta v_B$. The interfacial DMI can be described as an effective field along $x$ direction, which has the same magnitude but opposite signs for ↑↓ and ↓↑ DWs [1,2]. There are two parameters, the position X(t) and the conjugate momentum $\Phi$(t), describing DW motions, where $\Phi$(t) is defined as the in-plane angle with respect to the $+x$ direction. The corresponding equations without pinning field under $B_x$ are given by [2]

$$(1+\alpha^2)\frac{dX}{dt} = \frac{\gamma w_{dw} B_k}{2}\sin(2\Phi) - (1+\alpha\xi)u - \alpha\gamma w_{dw}\frac{\pi}{2}B_{SHE}\sin(\Phi) - \gamma w_{dw}\frac{\pi}{2}(B_x + B_{DMI})\cos(\Phi) \quad (1)$$

and

$$(1+\alpha^2)\frac{d\Phi}{dt} = -\frac{\alpha\gamma B_k}{2}\sin(2\Phi) - \frac{(\xi-\alpha)u}{w_{dw}} - \gamma\frac{\pi}{2}B_{SHE}\sin(\Phi) + \alpha\gamma\frac{\pi}{2}(B_x + B_{DMI})\cos(\Phi) \quad (2)$$

where $w_{dw}$ is the DW width, $B_k$ is the shape anisotropy field, $\alpha$ is the Gilbert damping, $\gamma$ is the gyromagnetic ratio, and $u = \mu_B PJ/eM_s$ is conventional adiabatic STT. $\xi$ is a dimensionless constant related with non-adiabatic STT, $J$ is the current density, $P$ is the spin polarization of the current, $e$ is the electron charge, $\mu_B$ is the Bohr magneton, and $M_s$ is the saturation magnetization. $B_{SHE} = \hbar\theta_{SH}J/2eM_s t_F$ is SHT, where $\hbar = h/2\pi$ and $h$ is the Planck constant, $\theta_{SH}$ is the spin Hall angle, and $t_F$ is the thickness of FM, and $B_{DMI}$ is the DMI effective field.

For the steady state and small values of $\Phi$, an analytical expression of DW velocity is given by

$$\frac{dX}{dt} \approx \pm\gamma w_{dw}\frac{\pi}{2}(B_x + B_{DMI})\frac{\frac{\pi}{2}B_{SHE}}{\frac{\pi}{2}B_{SHE} \pm \alpha B_k} - u\frac{\frac{\pi}{2}B_{SHE} \pm \xi B_k}{\frac{\pi}{2}B_{SHE} \pm \alpha B_k} \quad (3)$$

where +/- applies for $\Phi \approx 0$ or $\Phi \approx \pi$, respectively. Thus, $B_x$ induced velocity difference between ↑↓ and ↓↑ DWs can be written as follows:

$$\Delta v_{B_x} = \left|\gamma w_{dw}\frac{\pi}{2}(B_x - B_{DMI})\frac{\frac{\pi}{2}B_{SHE}}{\frac{\pi}{2}B_{SHE} + \alpha B_k} + \gamma w_{dw}\frac{\pi}{2}(B_x + B_{DMI})\frac{\frac{\pi}{2}B_{SHE}}{\frac{\pi}{2}B_{SHE} - \alpha B_k}\right| \quad (4)$$

where $B_{DMI}$ with the same magnitude but opposite signs are chosen for ↑↓ and ↓↑ DWs. We also ignore the conventional STT contribution to the velocity difference because no velocity difference was observed in previous works when $B_x$ = 0 mT [1,2]. Since $\frac{\pi}{2}B_{SHE} \approx \frac{\pi}{2}\times 48\text{mT} = 75\text{mT}$ is much large than $\alpha B_k = 0.1\times 15\text{mT} = 1.5\text{mT}$ by using $\mu_0 M_s = 1.3\text{T}$,

$\theta_{SH} = 0.1$, $J = 9 \times 10^7 \, \text{A cm}^{-2}$, $t_F = 0.6 \, \text{nm}$, and $\alpha = 0.1$ [1,2,17], $\Delta v_{B_x}$ is simplified to be $\pi \gamma w_{dw} B_x$. However, due to the pinning field and the instability of DWs like Walker breakdown in conventional STT driven DW motion [17,18], the velocity difference will have a saturated value of $\Delta v_{Max} = \pi \gamma w_{dw} B_{xth}$, where $B_{xth}$ is the threshold field at which $\Delta v_{B_x}$ reaches the saturation value. To reflect these conditions, $\Delta v_{B_x}$ can be written as

$$\Delta v_{B_x} = 2\gamma w_{dw} B_{xth} \text{Arctan}(\frac{\pi}{2} \frac{B_x}{B_{xth}}) \tag{5}$$

where $\Delta v_{B_x} = \pi \gamma w_{dw} B_x$ when $B_x \ll B_{xth}$, and $\Delta v_{B_x} = \Delta v_{Max} = \pi \gamma w_{dw} B_{xth}$ when $B_x \gg B_{xth}$.

By considering $B_x$ induced magnetization tilt,

$$R_H \propto (\eta_\uparrow - \eta_\downarrow)\cos(\theta_M) = (\eta_\uparrow - \eta_\downarrow)\sqrt{1-(\frac{B_x}{B_{kPMA}})^2} = 4\frac{\tau \gamma w_{dw} B_{xth}}{L_0} \text{Arctan}(\frac{\pi}{2} \frac{B_x}{B_{xth}})\sqrt{1-(\frac{B_x}{B_{kPMA}})^2} \tag{6}$$

where $\theta_M = \text{Arcsin}(M_x/M_s)$ is the magnetization tilt angle from $z$ axis and $B_{kPMA}$ is the perpendicular anisotropy field. This expression indicates that $R_H \propto B_x$ when $B_x$ is around zero and $R_H(B_x)$ does not depend on the current density because $\frac{\pi}{2} B_{SHE} \gg \alpha B_k$. It should be noted that it will be possible when $\Delta v_{B_x} = \Delta v_{Max}, |\eta_\uparrow - \eta_\downarrow| > 1$. This indicates that when the magnetization reaches its saturation value ($|\eta_\uparrow - \eta_\downarrow| = 1$), $\Delta v_{B_x}$ is still smaller than $\Delta v_{Max}$. In this case, Eq. (6) is still valid, but $B_{xth}$ should be the threshold field for magnetization reaching its saturation value. It should be mentioned that, when the magnetization reaches its saturation value, all the reversal domains are completely suppressed. This case happens when $2\Delta v_{B_x} \tau > L_0$, which indicates that the expanded domain areas are larger than the nucleated reversal domain areas in the same time interval. The solid lines shown in Fig. 3(a) are the fitting results by Eq. (6) using $B_{xth} = 23\text{mT}$ and $B_{kPMA} = 230\text{mT}$. Here, $B_{kPMA} = 230\text{mT}$ approaches that obtained by measuring $R_H$ as a function of $B_x$[19]. Because the magnetization can be completely reversed between two uniformly magnetized states in this case, $B_{xth} = 23\text{mT}$ is the threshold field that the magnetization reaches saturation. The good fitting results further demonstrate that this model can well describe the current induced perpendicular switching.

## V. Discussion and conclusion

We would like to point out that, although the sample dimensions in previous works were about 500 nm $\times$ 500 nm and a much short current pulse (~ 15 ns) was used, the domain nucleation and propagation process proposed here can still occur due to the narrow DWs (typical several nanometers) in these ultrathin films and a 6 ns rise time of the injected current pulse [3]. The dependence of $I_c$ on DW pinning strength also confirms our model [10]. All of these results indicate that this type of switching behavior is caused by current-induced domain nucleation and subsequent DW motion and is not resulting from coherent switching [3–5]. Because the domain nucleation in these structures can be assisted by other effects [11–14], the spin Hall angle calculated from $I_c$ by using coherent switching model may be overestimated [4,5].

In conclusion, we have proposed a magnetization reversal model for describing SO torque induced perpendicular switching. By measuring the current induced magnetization switching under different external fields, we demonstrate that this model can well explain the current induced perpendicular switching. This model indicates that this type of switching is not directly induced by the Rashba effect or SHE based on a coherent switching process, clarifying those debates on the reversal mechanism of SO torque induced perpendicular switching. On the other side, this model also provides an effective means to reliably control magnetic multidomain states in these structures.

**References:**

[1] S. Emori, U. Bauer, S.-M. Ahn, E. Martinez, and G.S.D. Beach, Nat. Mater. **12**, 611 (2013).

[2] K.-S. Ryu, L. Thomas, S.-H. Yang, and S. Parkin, Nat. Nanotechnol. **8**, 527 (2013).

[3] I.M. Miron, K. Garello, G. Gaudin, P.-J. Zermatten, M. V Costache, S. Auffret, S. Bandiera, B. Rodmacq, A. Schuhl, and P. Gambardella, Nature **476**, 189 (2011).

[4] L. Liu, C.-F. Pai, Y. Li, H.W. Tseng, D.C. Ralph, and R. a Buhrman, Science **336**, 555 (2012).

[5] L. Liu, O.J. Lee, T.J. Gudmundsen, D.C. Ralph, and R.A. Buhrman, Phys. Rev. Lett. **109**, 096602 (2012).

[6] C. Chappert, A. Fert, and F.N. Van Dau, Nat. Mater. **6**, 813 (2007).

[7] A. Moser, K. Takano, D.T. Margulies, M. Albrecht, Y. Sonobe, Y. Ikeda, S. Sun, and E.E. Fullerton, J. Phys. D. Appl. Phys. **35**, R157 (2002).

[8] J. Kim, J. Sinha, M. Hayashi, M. Yamanouchi, S. Fukami, T. Suzuki, S. Mitani, and H. Ohno, Nat. Mater. **12**, 240 (2013).

[9] K. Garello, I.M. Miron, C.O. Avci, F. Freimuth, Y. Mokrousov, S. Blügel, S. Auffret, O. Boulle, G. Gaudin, and P. Gambardella, Nat. Nanotechnol. **8**, 587 (2013).

[10] O.J. Lee, L.Q. Liu, C.F. Pai, Y. Li, H.W. Tseng, P.G. Gowtham, J.P. Park, D.C. Ralph, and R.A. Buhrman, Phys. Rev. B **89**, 024418 (2014).

[11] Y. Bazaliy, B. Jones, and S.-C. Zhang, Phys. Rev. B **57**, R3213 (1998).

[12] J. Fernández-Rossier, M. Braun, a. Núñez, and a. MacDonald, Phys. Rev. B **69**, 174412 (2004).

[13] Y. Nakatani, J. Shibata, G. Tatara, H. Kohno, A. Thiaville, and J. Miltat, Phys. Rev. B **77**, 014439 (2008).

[14] J. Shibata, G. Tatara, and H. Kohno, Phys. Rev. Lett. **94**, 076601 (2005).

[15] I.M. Miron, T. Moore, H. Szambolics, L.D. Buda-Prejbeanu, S. Auffret, B. Rodmacq, S. Pizzini, J. Vogel, M. Bonfim, A. Schuhl, and G. Gaudin, Nat. Mater. **10**, 419 (2011).

[16] T.A. Moore, I.M. Miron, G. Gaudin, G. Serret, S. Auffret, B. Rodmacq, A. Schuhl, S. Pizzini, J. Vogel, and M. Bonfim, Appl. Phys. Lett. **93**, 262504 (2008).

[17] E. Martinez, J. Phys. Condens. Matter **24**, 024206 (2012).

[18] T. Koyama, D. Chiba, K. Ueda, K. Kondou, H. Tanigawa, S. Fukami, T. Suzuki, N. Ohshima, N. Ishiwata, Y. Nakatani, K. Kobayashi, and T. Ono, Nat. Mater. **10**, 194 (2011).

[19] C. Bi, L. Huang, S. Long, Q. Liu, Z. Yao, L. Li, Z. Huo, L. Pan, and M. Liu, Appl. Phys. Lett. **105**, 022407 (2014).**Acknowledgments** We acknowledge support from MSTC (Nos. 2011CBA00602, 2011CB921804), NSFC (Nos. 61221004, 61274091, 61106119, 61106082), and the China Postdoctoral Science Foundation funded project (Nos. 2012M520421, 2013T60188).18